\documentclass[aps,pra,preprint,superscriptaddress]{revtex4}

\usepackage{graphicx}
\usepackage{amssymb}

\small
\begin{document}

\title{Mixing rules and the Casimir force between composite systems}

\author{R. Esquivel-Sirvent}
\email[Corresponding author. Email:]{raul@fisica.unam.mx}
\affiliation{Instituto de F\'{\i}sica, Universidad Nacional Aut\'onoma
 de M\'exico, Apartado Postal 20-364, D.F. 01000,  M\'exico}
 \author{George C. Schatz}
 \affiliation{Department of Chemistry, Northwestern University, Evanston, Il, 60208.}
 
 \date{\today}


\begin{abstract}
The Casimir-Lifshitz force is calculated between two inhomogeneous composite slabs, each made of a homogeneous matrix with spherical metallic inclusions.  The effective dielectric function of the slabs is calculated using several effective medium approximations and we compare the resulting forces as a function of slab separation and filling fraction.  We show that the choice of effective medium approximation is critical in making precise comparisons between theory and experiment. The role that the spectral representation of the effective medium plays in making a Wick rotation to the complex frequency axis is also discussed. 

\end{abstract}
\maketitle

\section{Introduction}
The prediction by H. B. G. Casimir \cite{Casimir48} that two neutral parallel plates made of a perfect conductor  will attract  each other due to zero point fluctuations  has gone from a theoretical curiosity to a force that has been measured with very high precision and that has implications in several fields.  Knowledge of Casimir forces  is needed to search  for deviations from Newtonian gravitation at short distances \cite{Mostepanenko08,Onofrio06,Onofrio11} and to detect chameleon fields \cite{Brax10}.   In  applied physics and engineering the Casimir force plays a role in micro and nano mechanical devices \cite{zhao04,Del05,barcenas05,esquivel06,gusso06,batra07,pinto08}.  

The generalization of the Casimir force to arbitrary materials
  was developed by Lifshitz \cite{Lif56,DLP} . In the Lifshitz theory the force  originates from
fluctuating electromagnetic fields in the macroscopic bodies. These fluctuating fields are related to the dielectric function of the materials via the fluctuation-dissipation theorem.  Although a closed form expression for the dispersive forces can be given only for parallel plates separated by a gap, Lifshitz theory is general since it includes both the retarded and non-retarded limits.  For perfect conductors Casimir's original result is recovered. 

The dependence of the Lifshitz formula  on the dielectric function and magnetic susceptibility \cite{note} can  be used to modify the behavior of the force. For example,  its sign can be changed by a suitable choice of the dielectric functions of the slabs and of the medium between them \cite{Parsegian}.  Also,  its magnitude can be changed using thermally activated structural phase transformations of materials like   AgInSbTe \cite{Torricelli10} or $TiO_2$ , external  magnetic fields \cite{Esquivel09,Esquivel10,Esquivel101} or by changing the charge carrier concentration of the slabs \cite{Chen07}.   

Composite materials provide additional issues in the evaluation of the Casimir force. Typical composites are obtained by combining two or more different materials such that at certain wavelengths the composite behaves as an homogeneous material with an effective dielectric function.  Composites such as $VO_2$ \cite{Esquivel02,Nori09}, aerogels \cite{aerogel}, composites with nickel particles \cite{Inui10} and most recently meta materials \cite{Irina,Dalvit1,Dalvit2,Yannopapas09,Zhao09} have been shown to change the Casimir force.  

In general,   a  composite material  is defined  as  a   mixture of $N$ different materials each with a physical property (such as a dielectric function) $\it  {p}_j$, each occupying a volume fraction $f_j$ such that $\sum_{j=1}^{N} f_j=1$. The problem then is to relate  the  average or effective value $\it{\tilde  p}$ to the component values.  In addition to dielectric function, other physical properties $\it{ p}_j$ of interest are the magnetic susceptibility,  their Lam\'e constants, the thermal conductivity etc. In this work we will be concerned only with the effective dielectric function $\tilde {\epsilon}$ which in general is frequency and wave vector dependent.

Understanding the effective properties of composites remains an important problem. Examples where this is an important issue include magnetodielectric photonic crystals \cite{FPerez09}, 
metamaterials \cite{Jin09}, surfaces and meta surfaces \cite{Alex10}.   Composite materials are also important in colloid chemistry, and there has been recent interest in the optical \cite{Reyes05} and  structural properties of such materials as a result of advances in the preparation of colloidal crystals and clusters \cite{Kalsin06}, and in the preparation of DNA-linked gold nanoparticle superlattice materials \cite{Lee10}.

How many ways are there to find the effective dielectric function of a composite system? In a recent paper by Prasad \cite{Prasad07} at least 12 different expressions or mixing formulas  for calculating the effective properties are presented and compared with  experimental results.  There is no general answer as to which is the best effective medium approximation (EMA). As stated by Grandquist and Hunderi, 
 the  merits of the various theories can be judged only when comparing with experimental data \cite{Grandquist78}.  Comparisons with more fundamental theories in which the properties of the constituent particles are explicitly included is also possible, as was applied for gold nanoparticles by Lazarides and Schatz \cite{Lazarides00}.

In this paper we calculate the Casimir-Lifshitz force between two composite slabs made of a dielectric  homogeneous matrix with spherical inclusions of Au. The effective dielectric function of the slabs is calculated using several of the effective medium approximations. Depending on the   choice of mixing formula,  the  resulting Casimir force changes significantly.

In the first section we present a brief summary of  the most common effective medium approximations. Also we discuss  the spectral representation and its importance in the proof of Kramers-Kronig relations needed to evaluate the effective dielectric along the imaginary frequency axis. Finally we present calculations of the Casimir force for different mixing formulas, showing the variations in force with the choice of mixing formula.

\section{Effective Medium Approximations}  

In this section we review some of the existing effective medium approximations without going through their derivation. The system we consider is an homogeneous matrix made of a material with a dielectric function $\epsilon_h$ (the host) which has inclusions of a second material with a dielectric function $\epsilon_i$ (inclusions).  The inclusions occupy a volume fraction $f$ of the matrix, also known as the filling fraction. The dependence on frequency of the dielectric functions will be suppressed. Without loss of generality we will assume spherical inclusions of radius $a$. 

The simplest approach used by Wiener \cite{Choy} was to take simple arithmetic and harmonic averages, that by definition will provide an upper and lower bound for $\tilde {\epsilon}$. This formula is
\begin{equation}
   \frac{\epsilon_i \epsilon_h}{f \epsilon_h+(1-f) \epsilon_i}\leq\tilde{\epsilon}\leq f\epsilon_i+(1-f)\epsilon_h .
\end{equation}

More rigorous bounds for the effective dielectric function were found by Hashin and Strickman \cite{Hashin62} using  variational procedures. Although the original derivation of these bounds was done for the magnetic susceptibility they are applicable to the dielectric function, electric conductivity and heat conductivities. 
The Hashin-Strickman bounds (H-S) are

\begin{equation}
\epsilon_h+\frac{\epsilon_h f}{\epsilon_h/(\epsilon_i-\epsilon_h)+(1-f)/3}\leq \tilde{\epsilon} \leq \epsilon_i+\frac{ (1-f) \epsilon_i}{\epsilon_i/(\epsilon_h-\epsilon_i)+f/3}.
\label{HS}
\end{equation}

 The fundamental property of the H-S bounds is that any EMA has to yield a dielectric function that falls between these two bounds.  It is important to notice that Eq. (1) makes no assumption on the shape of the inclusions. However the shape is important and the expression given in Eq. (2) is for spherical inclusions. 

The Maxwell-Garnett approximation is the most commonly used EMA cited in the text books due to its simplicity and physically intuitive derivation \cite{Choy}. 
The predicted dielectric function is 
\begin{equation}
\tilde{\epsilon}= \epsilon_h\frac{1+2 f \alpha}{1-f \alpha}
\label{MG}
\end{equation}
where $\alpha=\frac{\epsilon_i-\epsilon_h}{\epsilon_i+2\epsilon_h}$ is the polarizability of the inclusion (divided by $a^3$).
 
The Maxwell-Garnett approximation (MGA) assumes that the separation between the inclusions is large so that they act as independent scatterers and that the polarization effect on one of the particles due to its neighbors is instantaneous.  The first assumption limits the applicability of MGA to dilute systems with a low filling fraction $f$.    This assumption breaks down as the filling fraction increases and corrections to the Maxwell-Garnett theory have been proposed \cite{Grandquist77,Grandquist78,Noguez,Barrera}.  The limitations and range of validity of the Maxwell-Garnett theory have been discussed extensively by Ruppin \cite{Ruppin78}. 
  
   The Maxwell-Garnett theory  does not predict percolation transitions. For example if the host material is an insulator and the spherical inclusions are metallic, as we approach the random packing filling fraction an insulator-conductor transition can be expected.  An improvement to Maxwell-Garnett theory comes from Bruggeman \cite{brugg} who treated the two components of the composite in a symmetrical fashion.  
   In the Bruggeman approach the effective dielectric function is obtained by solving the equation:
   
\begin{equation}
f \left (\frac{\epsilon_i-\tilde{\epsilon}}{\epsilon_i+2 \tilde{\epsilon}} \right ) +(1-f)\left( \frac{\epsilon_h-\tilde{\epsilon}}{\epsilon_h+2 \tilde{\epsilon}} \right ) =0.
\label{bruggeman}
\end{equation}

Both Eq.(\ref{MG}) and Eq. (\ref{bruggeman}) are for spherical inclusions. However, they can be derived for other geometries. In the Appendix we present the Maxwell-Garnett and Bruggeman expressions for ellipsoids.  Also, both expressions can be generalized to nonlocal dielectric functions wherein the dielectric function depends on frequency and wave vector. \cite{Chang05}

As mentioned before,  there are many EMA expressions and not all of them will be considered in this paper. Besides the Maxwell-Garnett and the Bruggeman EMA we selected Looyenga's  approach \cite{Looyenga}, which is in good agreement with experimental measurements  of the effective optical properties of colloidal quantum dot systems \cite{Qiao10}. Looyenga's model for spherical inclusions assumes an equation for the effective dielectric function of the form
\begin{equation}
\tilde {\epsilon} =\left (  f \epsilon_i^{1/3}+(1-f)\epsilon_h^{1/3} \right)^3, 
\label{Looyenga}
\end{equation}
again assuming spherical inclusions. 
The models so far presented are a small sample of many of the mixing formulas that exist. We emphasize that there is no generally accepted best effective medium theory so it is important to study a broad range of theories to see what sensitivity there is to the computer Casimir force.

\subsection{Spectral representation and Rotation to the Imaginary axis} 

The effective dielectric function of a composite depends on the dielectric functions of the materials and on the geometry of the system, for example the shape of the inclusions.  These two effects can be separated using the so-called spectral representation developed independently by Fuchs \cite{Fuchs}, Bergman and Milton \cite{Bergman,Milton}.  
By defining a function
\begin{equation}
t=\frac{\epsilon_h}{\epsilon_h-\epsilon_i},
\end{equation}
that only has information on the material properties and a function $G(L)$, the spectral function, that contains information on the geometry of the system, the 
 effective dielectric function can be written 
\begin{equation}
\tilde{\epsilon}=\epsilon_h \left ( 1-f\int_0^1 \frac{G(L)}{t-L}dL \right)
\label{bergman}
\end{equation}
where $L\in[0,1]$ is the depolarization factor. The spectral function $G(L)$ can be interpreted as the density of geometrical or shape resonances \cite{Ghosh}. For example, for the Maxwell-Garnett model for spherical inclusions one has that $G(L)= \delta(L-(1-f)/3)$. 

Using the spectral representation one can derive the mixing formulas described in the previous section, including Winer bounds, Looyenga, Maxwell-Garnett and Bruggeman.  The spectral function (dielectric constant) of the components can be determined experimentally from reflectivity measurements \cite{Day00}, for example. 

There is a more fundamental consequence of the spectral representation. It can  be proven that any effective dielectric function obtained using the spectral representation satisfies the Kramer-Kronig relations \cite{Gorges95}.
The importance of the Kramers-Kronig relations in Casimir force calculations is that the Lifshitz formalism relies on a rotation of the frequency to the complex plane, that is  $\omega\rightarrow i\zeta$.  As shown in Ref. (\cite{Landau60}) rotation of the dielectric function is obtained from the Kramers-Kronig relation as  

\begin{equation}
\tilde { \varepsilon}(i\zeta)=1+\frac{2}{\pi}\int\limits_0^{\infty}d\omega
    \frac{\omega \tilde{\varepsilon}^{''}(\omega)}{\omega^2+\zeta^2}.
 \end{equation}

\subsection{Comparisons between the different models}

Consider a composite material that consists of a homogeneous host of $SiO_2$ and  spherical inclusions made of gold. The dielectric function of $SiO_2$ is described by an  oscillator model
\begin{equation}
\epsilon_h(i\zeta)=1+  \frac{C_{UV}}{1+(\zeta/\omega_{UV})^2}   +\frac{C_{IR}}{1+(\zeta/\omega_{IR})^2},
\label{sio2}
\end{equation}
where $C_{IR,UV}$ are the absorption strengths in the infrared and ultraviolet and $\omega_{IR,UV}$ the absorption frequencies \cite{Bergstrom97}. For the inclusions we consider spherical Au particles whose dielectric properties are obtained from tabulated data \cite{Palik} with a  low frequency extrapolation using a Drude model \cite{George}. 
  In figure (1) we show the dielectric function calculated as a function  of the filling fraction   for the Wiener bounds (upper and lower bound), Maxell-Garnett, Bruggeman, and Looyenga expressions.   
 The frequencies used $\zeta=0.02 \omega_p$ (top) and $\zeta=0.5\omega_p$ (bottom) were chosen arbitrarily.  For the top graph the effective dielectric function is plotted with a logarithmic scale. The large values of $\tilde{\epsilon}$  are due to the pole at zero in the Drude dielectric function that is responsible for the DC conductivity. 
At this frequency the dielectric function of Au dominates.   The Bruggeman model predicts a percolation transition at $f=0.33$ that is evident from the inflection in the plot.  
  For higher frequencies the dielectric function of Au decreases (and it should go asymptotically to 1 with increasing frequency) and the dielectric function of the host medium is comparable to that of Au. In this case, the Maxwell-Garnett, Looyenga and Bruggeman models give very similar results as seen in the bottom graph.

\section{Effective medium theories and Casimir forces}

To calculate the Casimir force we use the Lifshitz formula between two parallel slabs  with local dielectric functions $\epsilon_1(\omega)$ and $\epsilon_2(\omega)$ respectively,  separated by a gap length $L$. The gap is  filled with a dielectric function $\epsilon_3(\omega)$.
 Lifshitz result for the force per unit area is  
    \begin{equation} 
       \label{lifshitz}
       F=\frac{\hbar c }{2 \pi^{2}}\sum_{\nu=s,p}\int_{0}^{\infty} d\zeta \int_{0}^{\infty}dQ Q k_3 \frac{r_{13}^{\nu} r_{23}^{\nu}}{e^{2k_3 L}-r_{13}^{\nu} r_{23}^{\nu}},
       \end{equation}
       where $r_{ij}^{\nu}$ is the reflectivity between medium $i$ and $j$ for either $p$ or $s$ polarization, 
  $Q$ is the wave vector component along the
plates, $q=\zeta/c$ and $k_3=\sqrt{\epsilon_3 q^2+Q^2}$. The above expression is evaluated along the imaginary frequency axis $i\zeta$, and now the dielectric functions $\epsilon_i(i\zeta)$ ($i=1,2,3$) have to be considered. This expression is valid for $\hbar \omega > K_B T$, ($K_B$ is the Boltzman constant). 

In our case we assume $\epsilon_3=1$  and the dielectric functions for both slabs are the same, and are obtained from the effective dielectric function $\tilde \epsilon$. 
 The main contribution to the Casimir energy comes from frequencies in the vicinity of $\omega\sim c/2L$ 
or wavelengths of the order of $\lambda\sim 4 \pi L$. In  order to calculate the Casimir force between composite materials the wavelength has to be larger than the typical size of the inhomogeneities $a$. That is, the separation between the slabs and the size of the inhomogeneities has to satisfy $4 \pi L > a$. For example,  if we consider a composite made of a homogeneous matrix and spherical inclusions of radius $a=20$ $nm$, then the separations of the plates have to be larger than  $L>1.5$ $nm$.     

  The Casimir force per unit area predicted by Eq. (9) calculated for the different EMA is shown in Figure (2).   Using the common notation of the Casimir literature we have plotted the reduction factor $\eta$, which is  defined as the ratio of the force predicted using Eq. (9) to the force between perfect conductors ($F_0=-\hbar c \pi^2/240 L^4$).  The top panel of Figure (2) corresponds to a low filling fraction $f=0.015$ where the Maxwell-Garnet, Bruggeman and lower Wiener bound give a similar result. The difference between Bruggeman model  and the arithmetic average (upper Wiener bound) is of the order of $8\%$ when the plates are 100 $nm$ apart and $20\%$  for plate separation of 300 $nm$.  For a higher filling fraction of $f=0.25$ (bottom panel) the different models predict different values of the reduction factor, in particular at large separations.   Also,  for higher filling fractions  a bigger magnitude of the reduction factor is observed. This is  expected since the composite starts behaving more like a metal due to the higher fraction of Au.  
 Also, we see that the difference between the Bruggeman and arithmetic average results increases, being of the order of $20\%$ at 100 $nm$ and up to $62\%$ at 300 $nm$. These two models are singled out since they have been used in Casimir force calculations. 
   
   The difference between the different models as a function of filling fraction can also be seen in Fig. (3), where the reduction factor is plotted as a function of filling fraction. The separation between the plates is kept fixed at $L=100$$nm$.  For low filling fractions, the Bruggeman and Maxwell-Garnett models give the same results.  The difference between them as the filling fraction increases is due to the insulator-metal transition predicted by the Bruggeman model.  However, the difference with the arithmetic mean is significant for all filling fractions.

  \section{Conclusions}
 The use of composite media to modify the Casimir force requires a careful study of
the effective dielectric response of the system being used. As shown in this paper,
the value of the force is very sensitive to the choice of the effective medium
model.  In  the Casimir literature the upper Wiener bound and the Bruggeman approach
are commonly used, but as we have shown,  they can lead to very different
theoretical predictions for the Casimir force.   To  make precise comparison between
theory and experiments, in particular in high precision experiments, the choice of
model needs to be carefully considered  or misleading conclusions might arise.
Before any measurement or calculation  of the Casimir force is done, optical
experiments are needed to determine which is the most suitable effective medium
model for a particular set up.
 \section{Appendix. Ellipsoidal Inclusions}

If the inclusions are ellipsoids the Maxwell-Garnett and Bruggeman theories can be generalized in the following way. 
The polarizability tensor of the ellipsoidal particle can be written as
\begin{equation}
\alpha_{ii}=\frac{1}{3}\left ( \frac{\epsilon_1-\epsilon_h}{1+(\epsilon_1-\epsilon_h)L_i} \right ) abc,
\label{alphaell}
\end{equation}
where $a$,$b$ and $c$ are the lengths of the ellipsoid semi-axes and $L_i$ are the depolarization factors. For a sphere $L_i=1/3$ and for a prolate ellipsoid with $a>b=c$ we have
\begin{equation}
L_i=\frac{1-e^2}{2e^3}\left [ ln\left (\frac{1+e}{1-e}  \right) 2e\right ], 
\label{alphaellipsoid}
\end{equation}
where $e$ is the eccentricity of the elllipsoid. 
Replacing the previous expression in Eq.(3), the Maxwell-Garnett effective dielectric function for ellipsoids is obtained. 

Similarly, the Bruggeman formula can be generalized to ellipsoids as 
\begin{equation}
f \left (\frac{\epsilon_i-\tilde{\epsilon}}{\epsilon_i+(L_i^{-1}-1) \tilde{\epsilon}} \right ) +(1-f)\left( \frac{\epsilon_h-\tilde{\epsilon}}{\epsilon_h+(L_i^{-1}-1) \tilde{\epsilon}} \right ) =0.
\label{bruggeman2}
\end{equation}

\acknowledgements   R. E. S. acknowledges the hospitality of the Non Equilibrium Energy Research Center of Northwestern University  and the partial support of  CONACyT project no. 82474. This research was supported by the NERC EFRC of the US DOE (BES Award de-sc0000989). Helpful discussions with members of the ESF Casimir Network and L. W. Moch\'an are acknowledge.

\newpage
\begin{figure}[h]
 \begin{center}
\includegraphics[width=.9\textwidth]{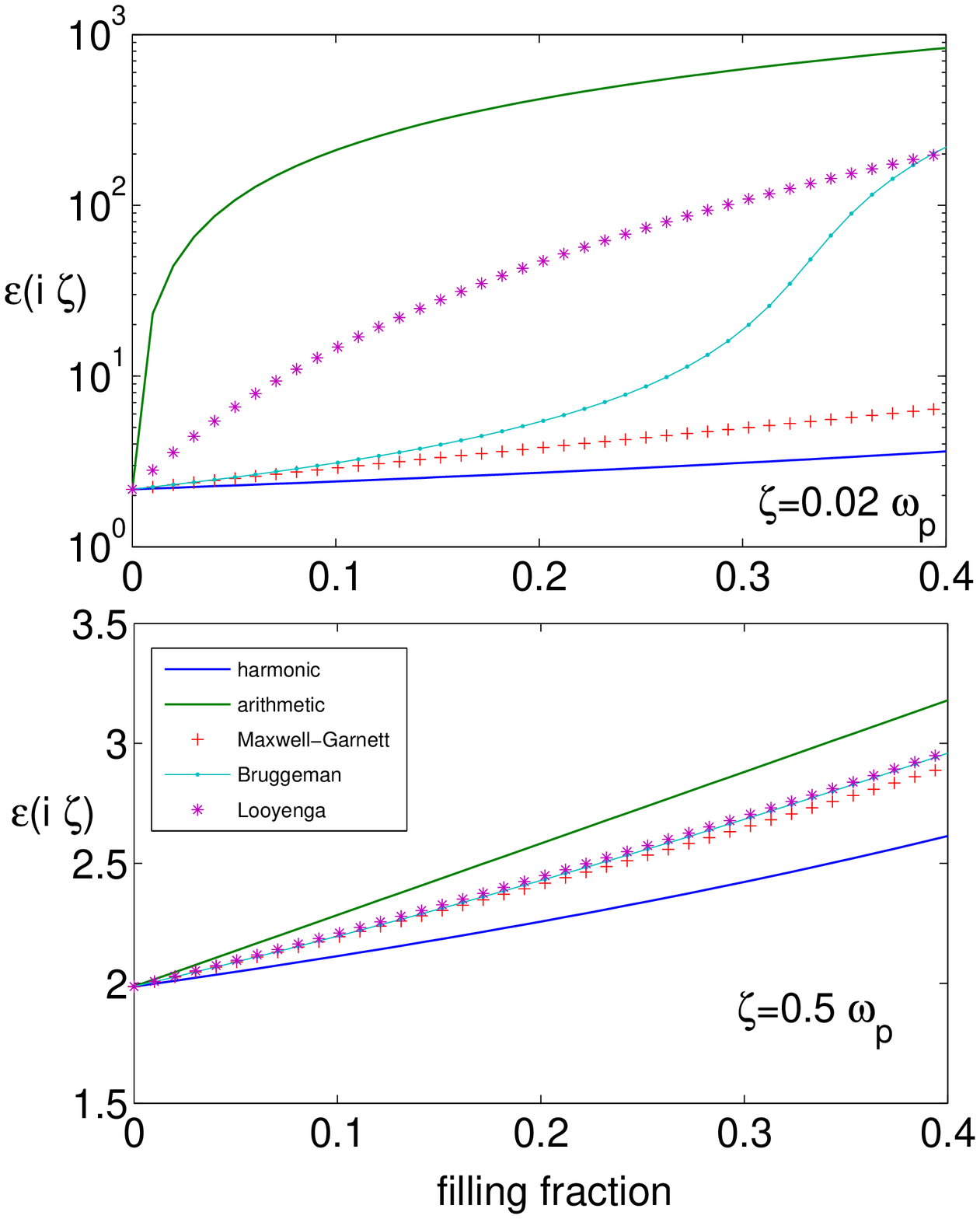}
\caption{ (Color online) Effective dielectric function plotted as a function of the filling fraction for a composite made of a $SiO_2$ host with  spherical Au inclusions.   In the top panel the effective  dielectric function is evaluated at the imaginary frequency  $\zeta=0.02 \omega_p$ and at $\zeta=0.5 \omega_p$  for the bottom graph.}
\end{center}
\end{figure}

 \begin{figure}[h]
 \begin{center}
\includegraphics[width=.9\textwidth]{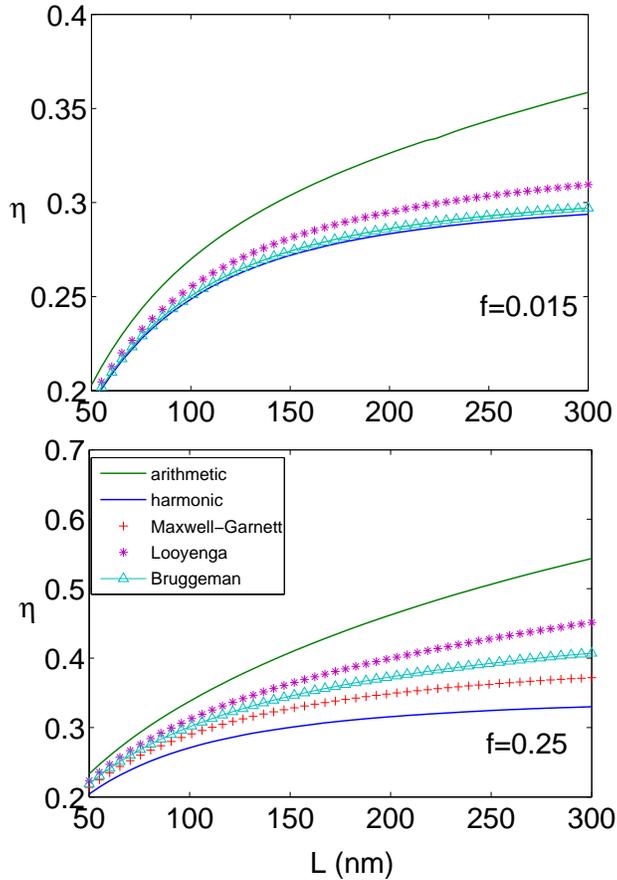}
\caption{(Color online)   Casimir force (normalized to the ideal case) as a function of plate separation between two composite plates made of a homogeneous matrix of $SiO_2$ with spherical inclusions of radius $a$. The results in this plot are valid provided that  condition  $4 \pi L > a$ is satisfied.  The top figure is for low filling fractions $f=0.015$. The Maxwell-Garnett and Bruggeman models give results similar to the lower Wiener bound.  For higher filling fraction ($f=0.25$) the difference in the force between the different models is much bigger in particular at large separations. } 
\end{center}
\end{figure}

\begin{figure}[h]
 \begin{center}
\includegraphics[width=.9\textwidth]{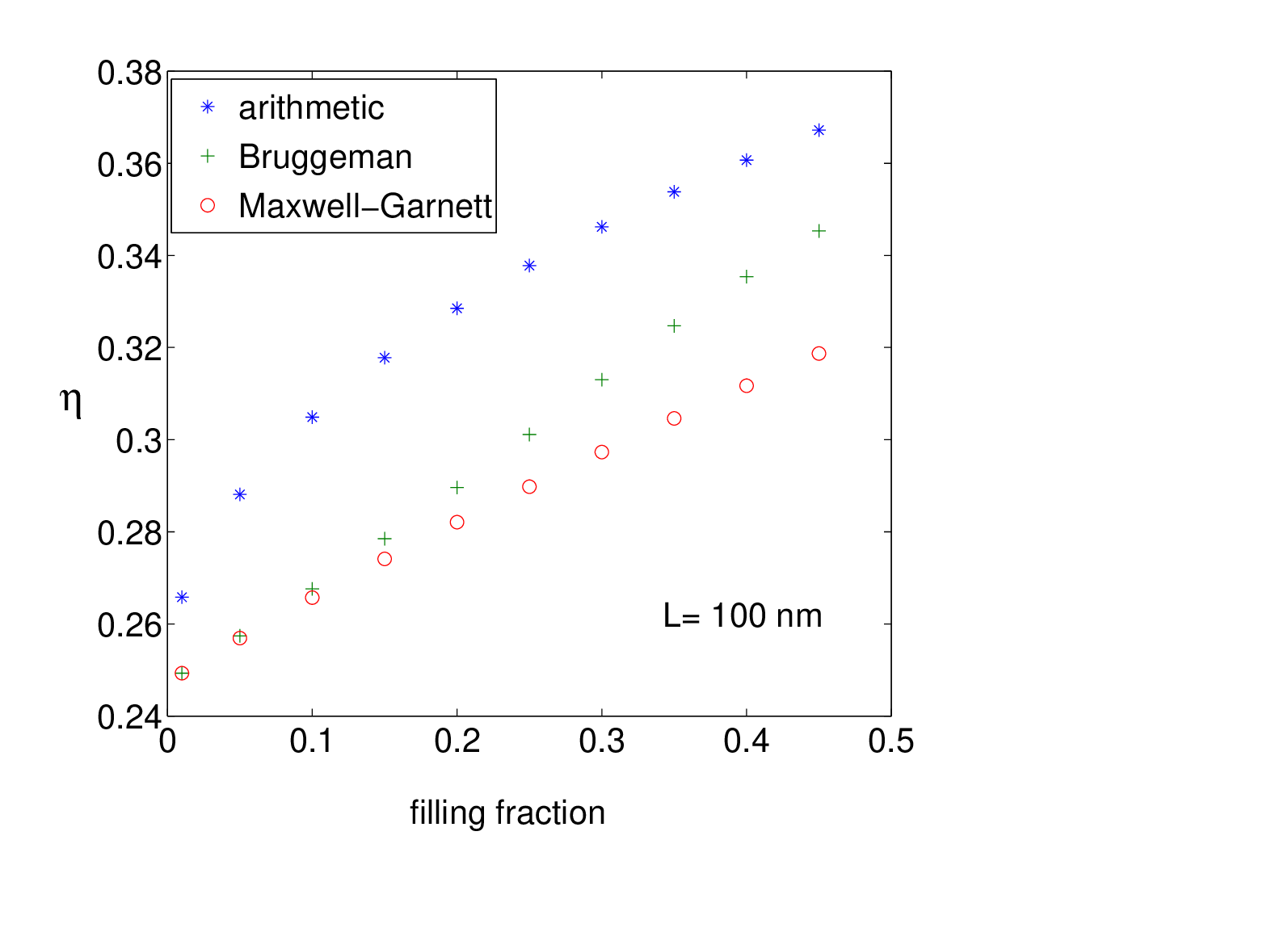}
\caption{ (Color online) Reduction factor as a function of filling fraction when the two plates are at  a fixed separation $L=100$$nm$.  At low filling fraction the difference between the Bruggeman and Maxwell-Garnet models is negligible but as the amount of Au particles increases,  the Bruggeman model correctly predict a more metallic behavior, increasing the force between the slabs.   } 
\end{center}
\end{figure}


\begin{thebibliography}{100}
\bibitem{note} In this work we assume a magnetic susceptibility of 1, that is true for most materials with the notable exception of metamaterials.
\end{thebibliography}
\end{document}